\documentclass[sigconf]{acmart}

\AtBeginDocument{%
  }

\copyrightyear{2025}
\acmYear{2025}
\setcopyright{acmlicensed}\acmConference[Websci '25]{Proceedings of the 17th ACM Web Science Conference 2025}{May 20--24, 2025}{New Brunswick, NJ, USA}
\acmBooktitle{Proceedings of the 17th ACM Web Science Conference 2025 (Websci '25), May 20--24, 2025, New Brunswick, NJ, USA}
\acmDOI{10.1145/3717867.3717900}
\acmISBN{979-8-4007-1483-2/2025/05}

\begin{document}

\title{Framing the Fray: \\ 
Conflict Framing in Indian Election News Coverage}

\author{Tejasvi Chebrolu}
\authornote{Work done while at UT Austin}
\email{tejasvi.chebrolu@research.iiit.ac.in}
\orcid{0009-0007-1441-0325}
\affiliation{%
  \institution{International Institute of Information Technology, Hyderabad}
  \city{Hyderabad}
  \state{Telangana}
  \country{India}
}

\author{Rohan Chowdhary}
\email{rohan.modepalle@research.iiit.ac.in}
\orcid{0009-0004-3061-9792}
\affiliation{%
  \institution{International Institute of Information Technology, Hyderabad}
  \city{Hyderabad}
  \state{Telangana}
  \country{India}
}
\author{N Harsha Vardhan}
\email{nemani.v@research.iiit.ac.in}
\orcid{0009-0000-6360-9110}
\affiliation{%
  \institution{International Institute of Information Technology, Hyderabad}
  \city{Hyderabad}
  \state{Telangana}
  \country{India}
}

\author{Ponnurangam Kumaraguru}
\email{pk.guru@iiit.ac.in}
\orcid{0000-0001-5082-2078}
\affiliation{%
  \institution{International Institute of Information Technology, Hyderabad}
  \city{Hyderabad}
  \state{Telangana}
  \country{India}
}

\author{Ashwin Rajadesingan}
\authornote{Corresponding author}
\email{arajades@austin.utexas.edu}
\orcid{0000-0001-5387-1350}
\affiliation{%
  \institution{University of Texas at Austin}
  \city{Austin}
  \state{Texas}
  \country{USA}
}

\renewcommand{\shortauthors}{Chebrolu et al.}

\begin{abstract}
  In covering elections, journalists often use conflict frames which depict events and issues as adversarial, often highlighting confrontations between opposing parties. Although conflict frames result in more citizen engagement, they may distract from substantive policy discussion. In this work, we analyze the use of conflict frames in online English-language news articles by seven major news outlets in the 2014 and 2019 Indian general elections. We find that the use of conflict frames is not linked to the news outlets' ideological biases but is associated with TV-based (rather than print-based) media. Further, the majority of news outlets do not exhibit ideological biases in portraying parties as aggressors or targets in articles with conflict frames. Finally, comparing news articles reporting on political speeches to their original speech transcripts, we find that, on average, news outlets tend to consistently report on attacks on the opposition party in the speeches but under-report on more substantive electoral issues covered in the speeches such as farmers' issues and infrastructure.
\end{abstract}



\maketitle

\section{Introduction}
News framing is a deliberate and strategic communication technique that involves carefully selecting specific aspects of an issue and amplifying their significance to effectively convey a message \cite{10.1111/j.1460-2466.1993.tb01304.x}. News frames in election contexts are especially consequential as they influence how people perceive this fundamental democratic exercise. For example, in news reporting, elections can be framed as a conflict to be won, emphasizing the competitive aspects of elections. Alternatively, they can be framed as a platform for exchanging ideas and debating issues, highlighting the importance of civic engagement and informed decision-making. 

In this paper, we analyzed the use of conflict frames in Indian election English-language news reporting. 
Conflict frames are defined as news frames that ``emphasize conflict between individuals, groups, or institutions as a means of capturing audience interest'' \cite{semetko2000framing}. Table \ref{tab:Conflict Examples} lists examples of article headlines with conflict frames. In the first example, Modi, leader of the Bharatiya Janata Party (part of the NDA coalition) and the current prime minister of India, attacks the Indian National Congress (part of the UPA coalition), the main opposition. Table \ref{tab:Conflict Examples} further highlights how various news outlets construct their conflict frames. For each outlet, we present two examples—one portraying the NDA as the aggressor and another depicting the UPA in that role. Notably, both the NDA and the UPA frequently engage in attacks on their opponents' character, yet the underlying reasons for these attacks are seldom mentioned. This pattern suggests a broader trend in political coverage, where conflict is emphasized over substantive discussions. Finally, we provide examples of headlines that do not employ conflict frames to offer a clearer contrast and better illustrate the concept

\begin{table*}[t]
\caption{Some examples of election news headlines are shown here. The first group shows headlines with conflict frames, including the aggressor (political coalition making the attack) and target (political coalition being attacked). The second group shows headlines without conflict frames. The source is shown for all headlines.}
\small
\setlength{\tabcolsep}{4pt}
\begin{tabular}{|p{0.65\textwidth}|p{0.07\textwidth}|p{0.06\textwidth}|p{0.155\textwidth}|}
\hline
\textbf{Conflict Frame Headline} & \textbf{Aggressor} & \textbf{Target} & \textbf{Source} \\
\hline
Media chasing Sadhvi Pragya must also talk about corruption against Congress, its allies: PM Modi & NDA & UPA & India Today \\
\hline
Rahul Gandhi mocks PM Modi's chowkidar campaign & UPA & NDA & India Today \\
\hline
Mamata accuses Modi of buying votes with black money converted to white through demonetisation & UPA & NDA & The Indian Express \\
\hline
Your dad was 'Mr Clean', but ended life as a 'bhrashtachari': PM Modi to Rahul Gandhi & NDA & UPA & The Indian Express \\
\hline
Vote for AAP to save India from Modi: CM Arvind Kejriwal & UPA & NDA & The Times of India \\
\hline
AAP, Cong jointly creating anarchy in Delhi: BJP & NDA & UPA & The Times of India \\
\hline
Rahul begins campaign with attack on BJP & UPA & NDA & The Hindu \\
\hline
BJP hits back at Rahul on personality-oriented politics & NDA & UPA & The Hindu \\
\hline
Rahul Gandhi compares Narendra Modi to Hitler & UPA & NDA & NDTV \\
\hline
Narendra Modi attacks Congress, Rahul Gandhi: His top 10 quotes & NDA & UPA & NDTV \\
\hline
Cong MLA uses derogatory word for BJP leaders at Rahul's rally & UPA & NDA & Zee News \\
\hline
Congress Fielding Two Batsmen to Take Blame For Poll Defeat: PM Modi & NDA & UPA & Zee News \\
\hline
Modi's 5 years 'most traumatic', should be shown exit door: Dr. Manmohan Singh & UPA & NDA & Republic World \\
\hline
Congress manifesto full of lies, hypocrisy: PM Modi & NDA & UPA & Republic World \\
\hline
\multicolumn{4}{|l|}{\textbf{Non-Conflict Frame Headline}} \\
\hline
Ramdev to urge people to vote for Modi & \multicolumn{2}{c|}{-} & The Times of India \\
\hline
Mamata Banerjee to begin poll campaign on Women's Day & \multicolumn{2}{c|}{-} & Republic World \\
\hline
L K Advani lauds Harsh Vardhan for pulse polio campaign & \multicolumn{2}{c|}{-} & The Indian Express \\
\hline
PM rules out third term, but makes robust defense of economic policies & \multicolumn{2}{c|}{-} & NDTV \\
\hline
AAP's former Delhi ministers to campaign in Varanasi, Amethi & \multicolumn{2}{c|}{-} & Zee News \\
\hline
Parties rope in celebs to jazz up campaigning in Himachal & \multicolumn{2}{c|}{-} & India Today \\
\hline
AAP government wins confidence vote & \multicolumn{2}{c|}{-} & The Hindu \\
\hline
\end{tabular}
\label{tab:Conflict Examples}
\end{table*}

We chose to focus on conflict frames as they are known to have mixed consequences for how people perceive democracy. On the one hand, conflict is inherent in politics, and reporting on such conflicts can highlight disagreement and differences, signalling to the citizens that they have political choices \cite{de2007conflict}. Further, conflict frames may engage and excite citizens, increasing voter turnout \cite{schuck2016s}. On the other hand, conflict frames, by emphasizing disagreement, which is often framed as attacks, increase polarization \cite{kim2020effects} and erode political trust \cite{mutz2005new}. Further, this way of framing distracts citizens from obtaining substantive policy or issue information, depressing political knowledge, and increasing cynicism \cite{cappella1997spiral}. Particularly in an increasingly hostile and polarized Indian political climate \cite{jaffrelot2021modi}, conflict frames may play a crucial role in how the electorate views their democratic systems and politics.

To study conflict frames, we fine-tuned a DistilBERT classifier \cite{distilbert} to identify conflict frames in news headlines from seven prominent Indian news outlets (N=69,400 articles) during the 2014 and 2019 Indian General elections. Using the classifier, we evaluated how these frames vary by media outlets' ideological bias, their primary modality (TV vs print), and which party they were reporting on. 

An enduring puzzle in framing research is understanding to what extent media coverage actually reflects the complex ground reality that it hopes to represent \cite{entman2007framing}. In our case, the question is whether news outlets overemphasize conflict over more substantive issues in their election coverage. The challenge in conducting such an analysis is that it is hard to separate out news reporting choosing to emphasize attacks between parties from the ground reality of political parties stepping up their attacks. We resolved this issue by constructing a unique dataset of campaign speeches and news articles that report on these speeches. By directly comparing the issues extracted from the transcript of campaign speeches to the issues covered in their corresponding articles, we were able to evaluate which issues were under and over-reported. 

We summarize the key findings of this study:
\begin{itemize}
    \item Media modality (TV-based vs print-based), rather than media bias, significantly correlates with the use of conflict frames in election news articles in India.
    \item Most sources do not exhibit ideological biases in portraying a party as the aggressor or target in articles with conflict frames. 
    \item Compared to regional parties, national parties are more prominently featured in headlines that employ conflict frames.
    \item Media coverage often under-reports on substantive issues such as farmer concerns and infrastructure in favour of highlighting attacks on the opposition when reporting on politicians' speeches.
\end{itemize}

\section{Related work}
\subsection{Conflict framing}
 Conflict frames are one of the most commonly used news frames \cite{semetko2000framing}. They highlight conflicts between individuals, groups, institutions, or countries. Conflict frames are \textit{generic} frames and are not related to a specific topic or issue \cite{de2005news}. These frames may highlight personal/substantive, civil/uncivil, deep/superficial or normative/factual aspects of political conflicts \cite{van2024online}. The use of conflict frames when reporting election-related news has mixed effects - erosion of political trust \cite{mutz2005new,cappella1997spiral} and decline in the approval of politicians \cite{forgette2006high} as well as increased awareness of the significance of political decision-making \cite{schuck2013explaining} and increased voter turnout \cite{min2004news}. Hence, understanding conflict framing in election news is crucial due to its complex effects on the electorate.

\subsection{(Over)emphasizing conflict frames}
Journalists and newsrooms strive to accurately reflect the happenings on the ground. However, with limited resources, journalists make deliberate choices on what events to report on and how to cover them, relying on heuristics such as novelty to determine coverage \cite{van2019mediatization}. These practices sometimes lead to coverage that often distorts reality \cite{entman2007framing}. There may be multiple factors that lead journalists to overemphasize conflict frames beyond the ground reality. Research suggests that providing a confrontational angle to news grabs attention and engages a larger audience \cite{mutz2015your}. Commercial interests \cite{mcmanus1994market} and rising competition \cite{bennett2016news} may also influence the use of conflict frames.

There is some evidence to suggest that the media may indeed overemphasize conflict frames over substantive issues. Through semi-structured interviews of journalists, \citet{bartholome2015manufacturing} found that journalists play an active role in the conflict frame-building process. They find that ``subtle methods of journalistic news production are applied to facilitate, emphasize, and sometimes even exaggerate conflict.'' In their interviews with journalists and content analysis of Swiss election campaign coverage, \citet{hanggli2010political} found that journalists do use conflict frames more than political actors. Yet, questions remain on the extent to which conflict frames are favoured over substantive issues. In this work, using a novel dataset of campaign speeches and their corresponding news reporting, we quantitatively uncovered how conflict frames may override other issue frames in reporting.

\subsection{News framing in online Indian news}
In recent times, online news media has played a crucial role in the Indian elections \cite{majo2019online}. However, there is little empirical work on online news media and Indian elections. Unsurprisingly, given the complex and well-funded party-affiliated `IT cells' \cite{chaturvedi2016troll}, the majority of research focuses on Narendra Modi and the BJP's use of social media in political campaigning \cite{pal2016twitter,pal2017mediatized}. \citet{neyazi2019channel} trace how the BJP's social media dominance often translates to significant coverage on traditional media. However, there is little empirical research on news framing in the Indian context (see \citet{jha2015media,doi:10.1177/1527476415575907} for exceptions). While conflict frames have not been explicitly studied, research suggests that ``the media environment was conducive to lively contestation'' during the recent election cycles, with media featuring parties often attacking each other \cite{neyazi20212019}. Our research aims to address the gap in existing literature, where there is a lack of large-scale analyses on how Indian news media employ conflict frames.

\subsection{News frame detection}
\textcolor{black}{Computational framing methods can be categorized into topic modelling, frequency modelling, neural network-based models, and cluster-based models \cite{ali-hassan-2022-survey}. Researchers have used topic modelling \cite{nguyen2015guided, topicmodels}, frequency modelling \cite{sanderink2020shattered}, and cluster-based models \cite{burscher2016frames} to analyze frames in large datasets. Embedding-based techniques enable cross-lingual frame projection \cite{field-etal-2018-framing}. Neural models, including LSTMs \cite{naderi} and RNNs, have also been used for frame detection. Transformer-based models like RoBERTa have been applied in multi-task learning scenarios \cite{cabot2020pragmatics}. Frame detection is often posed as a multi-label classification problem \cite{mendelsohn-etal-2021-modeling}, with pre-trained transformer models fine-tuned for this task \cite{khanehzar2019modeling}. Multilingual transfer learning has been used in low-resource contexts \cite{akyurek-etal-2020-multi}. Most existing neural network-based models draw inspiration from The Media Frames Corpus \cite{article}, which contains news articles discussing five policy issues: tobacco, immigration, same-sex marriage, gun control, and the death penalty. While these models perform well on their specific tasks, the frames they analyze and the contexts they address differ significantly from our focus. As a result these models and datasets are not particularly relevant to the Indian context. Therefore, we trained a new classifier with a new dataset tailored specifically to our research objectives.}

\section{Background: Indian electoral context}

In this work, we analyzed news articles published during the 2014 and 2019 Indian general elections. Through the general elections, candidates are elected to the Indian Parliament (Lok Sabha). The political party or coalition that secures the majority of seats in the Lok Sabha forms the government. In India, the United Progressive Alliance (UPA) and the National Democratic Alliance (NDA) are the two major political coalitions. The UPA is a center-left alliance led by the Indian National Congress (INC), while the NDA is a center-right alliance led by the Bharatiya Janata Party (BJP). The UPA was the incumbent alliance in the 2014 elections, while the NDA has been in power from 2019 and beyond (as of 2024).

In 2014, the polling dates spanned a period from 7 April to 12 May in nine phases, culminating in the declaration of results on 16 May. Similarly, in 2019, the polling dates extended from 11 April to 19 May in seven phases.

\section{Data}

\subsection{News articles}
    We analyzed news articles from seven major news outlets that are among India's most popular and influential English-language news media agencies \cite{Indian_Readership_Survey}. The Hindu, The Indian Express, and NDTV are considered either left or left-center (N=35,870 articles). While Zee News, The Times of India, Republic World and India Today are either right or right-center (N=33,530 articles). These biases are based on Media Bias Fact Check classifications (MBFC) \cite{Zandt_2024}. \textcolor{black}{MBFC uses stance on policy issues to determine ideological bias. Although US-based, its ratings on Indian sources align well with research showing that The Hindu, The Indian Express and NDTV provide an anti-(BJP) government frame while the other sources provide a pro-government frame in their coverage \cite{sen2022analysis,neyazi2020digital,garimella2024unraveling,bhat2020anti}.}
    
    \textcolor{black}{In contrast to the Global North, India's news landscape remains dominated by traditional print and television media. The online news outlets selected are relatively newer offshoots of established print or TV platforms, reflecting this media environment. We focused this study on nationally prominent, predominantly English-language outlets with diverse regional bases, analyzing full-length text articles and excluding any video material from both print and television-based media. These sources were chosen for their wide circulation, national presence, and regional diversity.
    Though not exhaustive, this selection provides a diverse sample of India's English-language national news outlets.}
    
    For each source, we collected the news articles published on their websites for the two general elections cycles. Our analysis is limited to these two election years as online archives were unavailable for previous election cycles for most news outlets. Republic World was launched in 2017, and hence, we analyze articles from this source only for the 2019 elections. Table \ref{tab:media} gives an overview of the dataset.
    
    We limited our data to the articles that were published between January 1st of that year until the day before the results were announced (1 January - 16 May 2014 and 1 January - 23 May 2019). Then, we use keyword-based matching on article headlines and URLs to identify election-related articles published during the analysis period. The list of keywords (such as election and vote) used is shown in the Appendix (Section \ref{Appendix Filtering}).

    \begin{table}[htbp]
    \caption{The number of articles belonging to a particular outlet, the outlet's bias (according to MBFC), and their modality.}
    \small
    \centering
    \renewcommand{\arraystretch}{1}
    \resizebox{\columnwidth}{!}{%
    \begin{tabular}{|c|c|c|c|}
        \hline
        \textbf{Source} & \textbf{Article Count} & \textbf{Bias} & \textbf{Modality} \\
        \hline
        The Hindu & 12,711 & Left-leaning & Print \\
        The Times of India & 15,158 & Right-leaning & Print \\
        The Indian Express & 13,748 & Left-leaning & Print \\
        NDTV & 9,411 & Left-leaning & TV \\
        India Today & 9,157 & Right-leaning & TV \\
        Republic World & 2,640 & Right-leaning & TV \\
        Zee News & 6,575 & Right-leaning & TV \\
        \hline
    \end{tabular}}
    \label{tab:media}
\end{table}
    
\subsection{Campaign speeches}

    We collected campaign speeches made by BJP and INC party leaders during election campaigning to evaluate how issues were reported by the news outlets (N=224 speeches). Some speeches were originally in Hindi and were translated with the aid of GPT4-Turbo \cite{openai-chatgpt-2023}. The translation prompt can be found in the Appendix (Section \ref{Appendix Translation}). We note that this dataset is not an exhaustive list of all speeches made during the campaigns, as some speeches were not recorded on the official websites. A random selection of five full-text speech translations was evaluated by an annotator fluent in both Hindi and English and was found to be accurately translated. The dataset and code used in this project are publicly available on GitHub. \footnote{\url{https://github.com/Ashwin-R/WebScience25-Conflict-Frames}}

\section{Methodology}

\subsection{Unit of analysis: Article headline}
\textcolor{black}{We analyzed conflict frames in news headlines because of their outsized influence on content interpretation and their high consumption. Headlines are ``the most potent framing device within the syntactical structure'' \cite{pan1993framing}, which influence article interpretation even when full text is read \cite{ecker2014effects}. Further, 6 out of 10 readers only read headlines \cite{boyer2014national}. Paywalls also increase headlines' importance as accessible information sources. This approach aligns with prior literature examining headlines for frame analysis \cite{chuey2024epistemic,akyurek-etal-2020-multi,liu2019detecting}.}

\subsection{Operationalizing conflict frame}
\textcolor{black}{We operationalized conflict frames in news headlines based on \citet{schuck2016s}'s criteria, which include references to multiple sides of an issue, mentions of conflict or disagreement, personal attacks between actors, or instances of reproach or blame. Given that conflict frames require at least two political entities, we automatically label headlines with fewer than two entities as \textit{not conflict}. For headlines containing two or more political entities, we train a classifier to determine the presence of a conflict frame based on the aforementioned criteria.}  

\subsection{Training data for conflict frame classifier}
\textcolor{black}{To create the training dataset for our classifier, we employed a two-stage annotation process. Initially, two human annotators classified 151 news headlines, each containing at least two political entities, into either conflict or non-conflict frames based on the aforementioned operationalization. A high inter-rater agreement was achieved (Cohen's $\kappa$ = 0.81). Subsequently, the annotators independently classified an additional 709 headlines with similar criteria, resulting in a total dataset of 860 headlines. The final dataset comprised 566 conflict frames and 294 non-conflict frames. This dataset was used for training our model to identify conflict frames in election news headlines.}

\subsection{Conflict frame classifier}

We finetuned a pre-trained \texttt{English DistilBERT} model\footnote{\url{https://huggingface.co/distilbert/distilbert-base-uncased-finetuned-sst-2-english}} for text classification using the 860 labelled samples. We performed a hyper-parameter search to optimize model performance using  \texttt{wandb \cite{wandb}}. Following established research \cite{zhang2020revisiting}, we extended the training epochs, as current literature suggests this approach improves both performance and stability for smaller datasets (<1000 samples). Our training approach follows a proven transfer learning methodology \cite{phang2018sentence}. We fine-tuned a pre-trained model, which was initially trained on language modeling and further fine-tuned on SST-2 for text classification prior to our task-specific training. This progressive learning approach enhances model robustness. The optimal hyperparameters are listed in the Appendix (Table \ref{tab:Hyperparameters}). The average values of the 5-fold cross-validation are summarized in Table \ref{Metrics}.

Applying the classifier to the full dataset of news headlines, we find that 11,225 of the 69,400 news headlines (19.29\%) have been classified as having a conflict frame. More classifier validation and robustness checks are mentioned in the Appendix (Section \ref{Appendix Masking}).

\begin{table}[htbp]
    \caption{5-fold cross-validation metrics of the classifier used to identify frames in election news headlines.}
    \centering
    \begin{tabular}{|c|c|c|c|}
        \hline
        \textbf{Class} & \textbf{Precision} & \textbf{Recall} & \textbf{F1-Score} \\
        \hline
        Conflict & 0.92 & 0.91 & 0.92 \\
        Non-Conflict & 0.81 & 0.82 & 0.81 \\
        \hline
        Macro average & 0.87 & 0.87 & 0.87 \\
        \hline
    \end{tabular}
    \label{Metrics}
\end{table}

\subsection{Extracting political parties and politicians}
\label{political parties}

\textcolor{black}{We identified politicians and political parties mentioned in article headlines by matching with data from official party websites and the LokDhaba Indian Elections dataset \cite{kumar2020lokdhaba}. Then, we mapped mentions of politicians to their respective parties. In total, 47,376 articles mentioned at least one party in their headlines, with 16,827 mentioning at least two parties.} Further details and a worked example detailing the matching process are mentioned in the Appendix (Section \ref{Appendix Parties}).

\begin{table*}[!htbp] 
\caption{Results from random-effects logistic regressions modeling conflict frames in the headlines as the dependent variable, including random effects for news source, election year control variable, and other variables of interest.}
  \centering   
\small 

\setlength{\tabcolsep}{6pt} 
\renewcommand{\arraystretch}{1.2} 
\begin{tabular}{@{}lccc@{}} 
\hline 
\hline 
 & \multicolumn{3}{c}{\textit{Dependent variable:}} \\ 
\cline{2-4} 
 & \multicolumn{3}{c}{Is conflict frame} \\ 
 & Section 6.1 & Section 6.1 & Section 6.2\\
 & (Media bias) & (Media modality) & (Regional vs national party)\\ 
\hline 
 Mentions national party &  &  & 3.277$^{***}$ \\ 
  &  &  & (0.046) \\ 
 Mentions regional party &  &   & 2.085$^{***}$ \\ 
  &  &  & (0.027) \\ 
 Media bias: left-leaning (vs right-leaning) & $-$0.228 & 0.068 & 0.090 \\ 
  & (0.256) & (0.175) & (0.094) \\ 
 Primary modality: TV-based (vs print-based) &  & 0.707$^{***}$ & 0.467$^{***}$ \\ 
  &  & (0.180) & (0.094) \\ 
 Election cycle: 2019 (vs 2014) & 0.304$^{***}$ & 0.304$^{***}$& 0.259$^{***}$ \\ 
  & (0.023) & (0.023) & (0.028) \\ 
 Constant & $-$1.657$^{***}$ & $-$2.188$^{***}$ & $-$5.271$^{***}$ \\ 
  & (0.172) & (0.171) & (0.102) \\ 
\hline \\[-1.8ex] 
Observations & 69,400 & 69,400 & 47,376 \\ 
Log Likelihood & $-$29,957.160 & $-$29,953.050 & $-$20,734.490 \\ 
\hline 
\hline 
\multicolumn{4}{r}{\textit{Note:} $^{*}$p$<$0.1; $^{**}$p$<$0.05; $^{***}$p$<$0.01} \\ 
\end{tabular}

\label{Regression Table}
\end{table*}
\subsection{Extracting aggressors and targets}

\textcolor{black}{For headlines that employed conflict frame, we identified the aggressors and targets by analyzing the structure of the headline. Using GPT-4 Turbo \cite{openai-chatgpt-2023}, we extracted these roles, typically with the aggressor as the speaker or subject and the target as the object of the headline respectively. Then, we mapped politicians to their respective political parties using the approach described above. This method was validated with a list of 100 headlines, human-annotated for aggressors and targets, achieving 98\% accuracy. Accuracy was calculated as the percentage of headlines where both the aggressor and target were correctly identified. Further metrics and additional details can be found in the Appendix (Section \ref{Appendix Agrressors}}).

\subsection{Matching campaign speeches to news articles}

\textcolor{black}{For each speech in the speech dataset, we manually extracted the location and speaker from the titles of the speech, discarding unclear cases. Articles were matched to speeches if published within three days of the speech, mentioning the speaker, location, and at least one speech-related term in the article text. This approach matched 145 speeches to 607 articles. A manual evaluation of 50 samples showed 85\% accuracy. Accuracy is defined as the percentage of matched articles that actually mentioned or discussed the speech in question. The full list of speech event terms (such as `rally' and `meeting') is shown in the Appendix (Section \ref{Appendix Rally Keywords}).}

\subsection{Extracting electoral issues mentioned in speeches and news articles} \label{electoral_issues}

\textcolor{black}{To identify electoral issues, we used data from the \href{https://www.lokniti.org/media/PDF-upload/1536927390_2768500_download_report.pdf}{2014 (Q8)} and \href{https://www.lokniti.org/media/PDF-upload/1570173782_98991600_download_report.pdf}{2019 (Q12, Q38)}  National Election Study Pre-Poll Surveys which surveyed participants on the importance of different issues. We selected issues that at least 1\% of the survey respondents rated as important. This included 19 issues, such as healthcare and electricity. We also included an `Opposition' issue, which corresponds to mentions of the opposition party or its members. To identify these mentions of the opposition, we use the approach described earlier (Section \ref{political parties}). These opposition party mentions in campaign speeches are invariably negative and attacking in nature.}

\textcolor{black}{To identify which of these issues were covered in the speeches, we followed \citet{muddiman2017news}'s dictionary-based approach. Initially, we generated a list of the most frequently used 5,000 stemmed terms from speeches and articles after removing stop words. Then, in the first pass, three annotators manually scanned through these terms and mapped all relevant terms to each issue based on their reading of party manifestos and domain knowledge. For each candidate term, 10 sentences containing the term were sampled and manually labelled for topic relevance (1 or 0). For the first 100 candidate terms (1000 sentences), three annotators achieved high inter-rater reliability in their topic relevance coding (Fleiss $\kappa = 0.84$). Then, the three annotators independently labelled the remaining terms. The candidate terms were removed if less than 80\% of the sampled sentences related to the issue.  This process yielded a dataset of 19 issues containing 139 keywords from 584 candidate terms. The list of issues and keywords can be found in the Appendix (Table \ref{tab:categories_keywords}).}

\section{Analysis}

The regression analyses were performed using the $lme4$ R package \cite{bates2015fitting}. The marginal means estimation and the planned contrasts were conducted using the $emmeans$ R package \cite{lenth2022emmeans}.

\subsection{How do media bias and modality relate to media outlets' use of conflict frames?}

We conducted a random-effects logistic regression on the 69,400 news headlines dataset, modeling the presence of a conflict frame in the news headline as a dependent variable (0 or 1) with a random effect for news source, a control variable indicating election year (2014 or 2019) and the variable of interest, media bias (left or right). To evaluate how the modality of the source relates to the use of conflict frames, we used an identical logistic regression model with an additional independent variable indicating whether the source was print-based or TV-based. The results of the two regressions are available in the left and center columns of
Table \ref{Regression Table}.

From the left column in Table \ref{Regression Table}, we find that the coefficient for the media bias indicator is not statistically significant. Thus, we did not find reliable evidence of an association between the ideological bias of the media outlet and the frequency with which it employs conflict frames in election news. Whereas, from the center column, based on the co-efficient for the primary modality indicator, we find that the odds of TV-based news outlets employing conflict frames is twice the odds of such usage at print-based news outlets ($b=0.707, OR = 2.030, SE = 0.17982, z$-$score = 3.933, p < 0.001$). This difference is statistically significant. These results suggest that in English-medium Indian election news, conflict frames are driven more by the media modality (TV vs. print) than by ideological bias.  

\subsection{How does the use of conflict frames vary when reporting on national and regional parties?}

We conducted a similar logistic regression as before and included two indicator variables indicating the presence of national and regional parties. Note that we use separate indicator variables as the headlines may include both national and regional parties. For this analysis, since we were comparing the two party types, we only included articles that mention at least one party (n  = 47,376).

From the coefficient for the national party indicator in the right column in Table \ref{Regression Table}, we find that the mention of a national party in the headline is positively associated with the headline using a conflict frame ($b = 3.277, OR = 26.490, z$-$score = 71.810, p < 0.01$). Similarly, the presence of a regional party is also positively associated with the headline using a conflict frame \texttt{($b = 2.085,OR = 8.047, z$-$score = 77.129, p < 0.01$)}. Further, the odds of using conflict frames in headlines mentioning national parties is more than three times that of headlines mentioning regional parties. This difference is statistically significant (via Wald test, \texttt{W(1) = 745.61, $p < 0.01$}).

\begin{figure*}[t]
    \centering
    \includegraphics[width=\textwidth]{figures/combined_err_plot.png}
    \caption{95\% CI of probabilities for major political coalitions as aggressors (a) and targets (b) in conflict frame headlines across media outlets. Left-leaning media and UPA are shown in blue; right-leaning media and NDA are in orange. The CI values and estimates are detailed in Table \ref{tab:alliance_attacked} and Table \ref{tab:alliance_being_attacked}.}
    \label{fig:coalition}
\end{figure*}

\subsection{How do news outlets differ in reporting which parties and coalitions are attacking or being attacked in conflict frame headlines?} 

We aggregated by source, the proportion of conflict frame articles that portray the UPA (left-wing coalition including the INC) and the NDA (right-wing coalition including the BJP) as attacking another party or being attacked by another party. Figure \ref{fig:coalition} shows, aggregated by the news outlet, the 95\% CI of the proportion of conflict headlines where the coalitions are the aggressors (a) and where the coalitions are the targets (b). 

From Figure \ref{fig:coalition} (a), we observe that a majority of sources (India Today, The Times of India, The Indian Express, NDTV) across both years report on similar proportions of conflict-framed headlines where NDA and UPA coalitions are the aggressors. The differences between the proportion of UPA attacking ($a\textsubscript{UPA}$) and NDA attacking ($a\textsubscript{NDA}$) headlines are statistically significant for The Hindu ($a\textsubscript{UPA}= 0.34,a\textsubscript{NDA}= 0.29, z$-$score = 2.231, p < 0.05$), Republic World ($a\textsubscript{UPA} = 0.31, a\textsubscript{NDA}=0.37, z$-$score = -2.729, p < 0.01$), and Zee News ($a\textsubscript{UPA} = 0.31, a\textsubscript{NDA}=0.39, z$-$score = -4.690, p < 0.01$) in 2019. The Hindu (left-leaning) published a higher proportion of headlines with the UPA coalition (left-leaning) as the aggressor, while the Republic World and Zee News (both right-leaning) published a higher proportion of the NDA coalition (right-leaning) as aggressors, which was consistent with their ideological biases.

From Figure \ref{fig:coalition} (b), most sources across both years report a significantly higher proportion of conflict-framed headlines where the NDA is being attacked compared to the UPA. Zee News in 2014 and Republic World in 2019 do not exhibit a statistically significant difference in their reporting of NDA and UPA being attacked. \footnote{\textcolor{black}{We do not print the proportion percentages in the figure for readability reasons. Appendix Tables \ref{tab:alliance_attacked} and \ref{tab:alliance_being_attacked} contain all the proportion percentages.}} 
Somewhat surprisingly, even in the case of the Republic World, known for its extreme pro-BJP stance \cite{garimella2024unraveling}, over 40\% of news headlines with conflict frames portray the NDA as being attacked.

\subsection{Are conflicts over-emphasized compared to electoral issues?}
To evaluate this question, we compared the issues mentioned in the speeches to the issues mentioned in the news reporting on the speeches using the issue keywords identified in Section \ref{electoral_issues}. First, the headlines of news articles reporting on campaign speeches were classified as having a conflict frame at a much higher rate than the full dataset (\texttt{31.96\% vs 19.29\%, z-score=10.48, $p <0.01$}), suggesting that campaign speeches are disproportionately reported with a conflict frame. In contrast, only 1.1\% of news articles reporting on campaign speeches included a headline mentioning one of the issue topics, highlighting the significant underreporting of issues compared to conflicts.

\begin{figure}[h]
        \centering
        \includegraphics[width=0.475\textwidth]{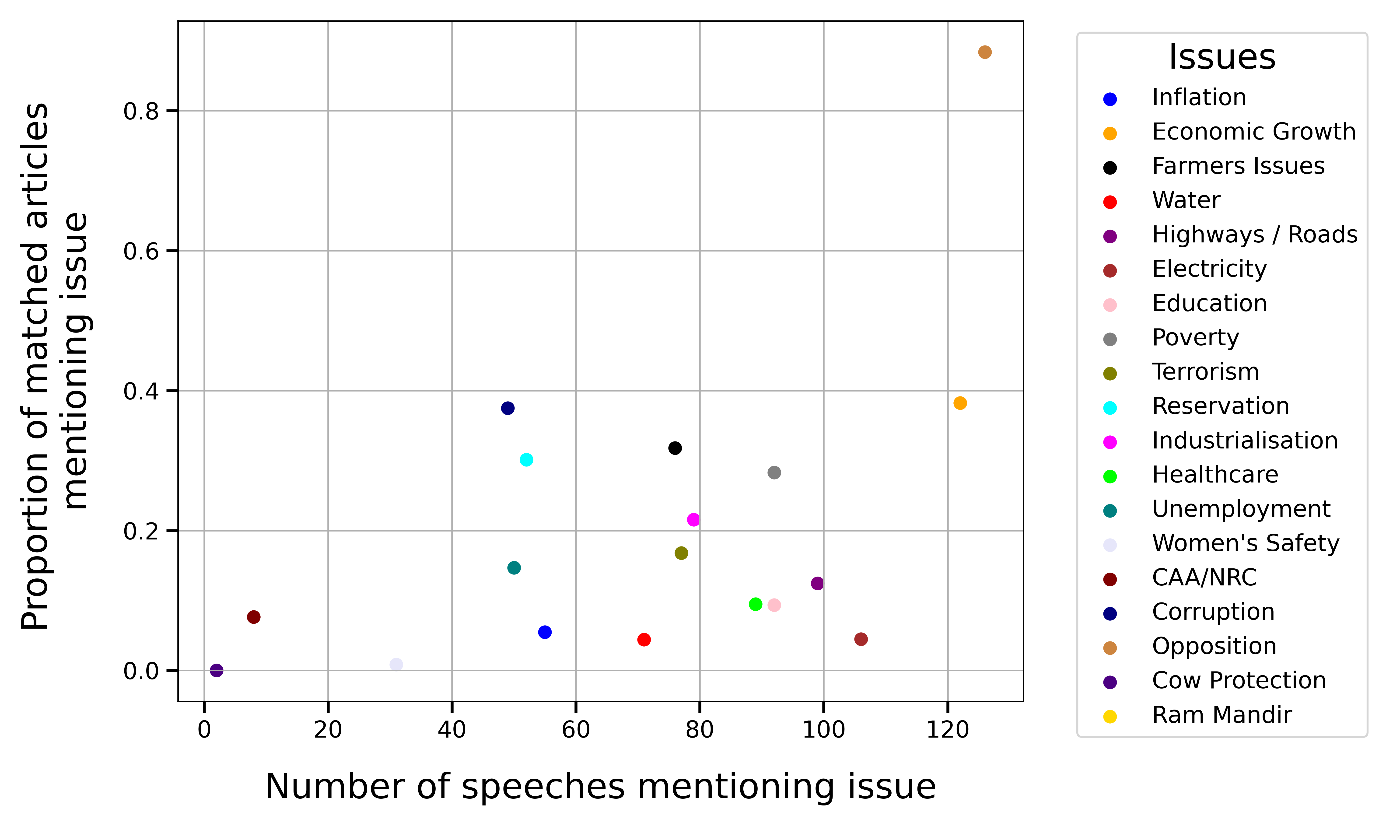}
        \caption{The frequency of speeches mentioning a specific issue versus the proportion of articles mentioning the same issue. Analysis shows substantive topics, like farmers' issues, are underrepresented (bottom-left), whereas the most frequently mentioned issue is the opposition. The frequency and proportion values are detailed in Table \ref{tab:issues_table}.}
        \label{fig:Topic vs Likelihood}
    \end{figure}

Examining the full content of the matched news articles, Figure \ref{fig:Topic vs Likelihood} shows a scatter plot with the x-axis representing the number of speeches that mention an issue and the y-axis representing the proportion of matched articles that mention that issue. Higher y-values imply that a higher proportion of articles that were matched to the speech discuss that issue. 

    From Figure \ref{fig:Topic vs Likelihood}, we find that most speeches reference the opposition, likely with attacking intent and given that the speech mentions the opposition, 88\% of articles that report on the speech also mention the opposition in the article text. However, this proportion drops significantly when speeches mention other more substantive issues. For example, only about 30\% of articles reporting on campaign speeches that mention farmer issues actually mention farmer issues. Thus, while news outlets do not technically over-report on conflict (since political campaign speeches more often than not deride the opposition party), they appear to underreport more substantive policy issues.

\section{Discussion}

\subsection{Effect of media ideological bias in online news}

From our analyses, ideological biases of the news outlets were not significantly associated with the use of conflict frames. Barring some exceptions in the 2019 election cycle, we found that sources, irrespective of their ideological biases, did not significantly differ in terms of the proportion of news headlines that portrayed the UPA and NDA coalitions as aggressors. Surprisingly, given the relative lack of press freedom\footnote{\url{https://rsf.org/en/country/india}}, most sources, irrespective of their ideological biases, published a much higher proportion of news headlines where the right-leaning NDA was targeted. These results suggest that ideological bias doesn't correlate with conflict framing choices in Indian election news coverage.

Other factors may play a more substantial role in selecting conflict frames. We find that TV-based news outlets publish a higher proportion of conflict-frame headlines than print-based outlets. This finding is in contrast to past studies in Europe, which suggest that the major difference was not between print and TV media but between sensationalist and sober media \cite{semetko2000framing}. Further, other research suggests that conflict frames increase readership, and hence, media outlets may have commercial interests in publishing news headlines with conflict frames \cite{mcmanus1994market}. Also, interestingly, even Republic World, the heavily pro-BJP/NDA news outlet \cite{garimella2024unraveling} published over 40\% of conflict frame headlines targeting the right-wing NDA coalition. One potential reason is that attacks on the in-party can provoke anger and mobilize partisans \cite{gervais2019rousing}, leading to increased consumption of partisan content. Further, conflict frames highlight disagreements and contentious interactions and are especially effective at drawing viewers’ attention \cite{semetko2000framing}. Since higher viewers translate into increased advertising revenue, outlets have a strong incentive to use conflict frames, even if the attacks are against the coalition they may be biased towards, in order to boost viewership and, ultimately, their profits \cite{mcmanus1994market}. An alternate reason could be that the media reports simply reflect the inter-party dynamics on the ground. Since NDA coalition has emerged as the stronger coalition over the past decade, winning three consecutive elections, it is likely that other parties have stepped up their attacks on the NDA resulting in an uptick of such reporting.

\subsection{Pathway to  political polarization}

Our findings indicate that news outlets use conflict frames more frequently when reporting on national parties than regional parties. As national parties are followed by more citizens compared to regional parties, this may further contribute to public perception that politics is highly negative and hyperpartisan, exacerbating polarization. Research suggests that citizens draw their cues from political elites and respond to elite polarization by expressing negative evaluations of the outparty, resulting in what political scientists call affective polarization \cite{banda2018elite}. Affective polarization has significant negative consequences, such as reducing political trust and lowering support for bipartisanship \cite{kingzette2021affective} (see \cite{broockman2023does} for an alternate perspective).

Comparing campaign speeches and the news articles that reported on them, we find that news articles consistently report on mentions of the opposition party in the speeches, which are likely negative in nature, while significantly under-reporting on electorally consequential issues such as farmer welfare, which were also brought up in the speeches. Such emphasis on conflict rather than substantive issues may result in lower knowledge about political issues among citizens \cite{cappella1997spiral} while exacerbating polarization \cite{kim2020effects}. It can also lead to an erosion of public trust \cite{mutz2005new,cappella1997spiral}, and decline in the approval of politicians \cite{forgette2006high}.

\section{Limitations and Future Work}
While our research provides meaningful insights, it is not without limitations. Our analysis is restricted to conflict frames. Although conflict framing is critical to understanding election news reporting, incorporating other frames, such as strategy or game frames, could yield alternative perspectives. Future research may explore these additional frames, identify multiple frames, and integrate multimedia content such as images to develop a better understanding of election news framing. Similarly, while the regression analyses performed were useful, we likely did not control for other potential confounding factors influencing conflict framing such as media ownership or market share. The dataset includes articles from only English-language news outlets, which, while widely read, exclude regional news outlets and Indian-language publications. It is unclear if these results can be generalized to regional language publications. Expanding the dataset to include diverse sources could provide a more representative analysis.

This work focuses primarily on headlines, given their outsized influence on readers' perceptions. However, the body of the articles and multimedia elements, may also play significant roles in framing and warrant further investigation. The campaign speech dataset is constrained to those available on the official websites of national parties (INC and BJP). Regional parties often lack archived speeches, limiting the scope of our analysis. Finally, the research is contextually focused on India, precluding cross-country comparisons. A comparative analysis with other democratic systems could provide valuable insights into the universality or uniqueness of conflict framing in election coverage. Additionally, analyzing differences in headline framing across media outlets with varying political orientations could also provide more details on how media bias could influence conflict framing.

\section{Ethical Considerations}

\textcolor{black}{Studying media processes in India is crucial, given the complex and evolving landscape of press freedom in the country. There have been significant challenges to press freedom in the country with \textit{Reporters Without Borders} ranking India at 159 out of 180 countries in their Press Freedom Index\footnote{\url{https://rsf.org/en/index}}. In this context, we aimed to provide a nuanced perspective of how news outlets employ conflict frames while acknowledging that the political climate has seen increasing pressure on media outlets and journalists in recent years. All analyses presented in the paper are at the news source level, do not identify individual reporters and use publicly available news articles.}

\begin{acks}
The authors acknowledge the generous support from the Center for Media Engagement and the Moody College of Communication at The University of Texas at Austin. They also acknowledge various students at Precog, IIITH for valuable suggestions which helped the paper immensely.
\end{acks}

\bibliographystyle{ACM-Reference-Format}
\bibliography{main}

\appendix
\label{sec:appendix}

\section{Keywords used for filtering articles}
\label{Appendix Filtering}
The keywords used to filter for articles and decide whether a given article was election-based or not are as follows:
\\\\
\textit{Campaign, rally, campaigning, rallying, Poll, poll, polling, Election, election, Voting, votes, vote, voter, EC, EVM, EVMs, Manifesto, Lok Sabha, Parliamentary, electoral, lok-sabha, INC, Congress, BJP, TDP, Bharatiya Janata Party, Telugu Desam Party, YSRCP, TMC, Trinamool Congress, TRS, Telangana Rashtra Samithi, BSP, Bahujan Samaj Party, SP, Samajwadi Party, AAP, Aam Aadmi Party, NCP, Nationalist Congress Party, DMK, Dravida Munnetra Kazhagam, AIADMK, All India Anna Dravida Munnetra Kazhagam, Biju Janata Dal, BJD}.

\section{Translation prompt}
\label{Appendix Translation}
The following prompt was used to translate speeches from Hindi to English:
\small
\begin{verbatim}
Given some content in Hindi, translate it 
to English while retaining the original 
meaning.
Return the translated content. If 
there is English content in the input, 
keep the English content as is.
The output should be in the 
following format:
{
    'translated_content': 
    'Translated content here'
}
Translate the following Hindi content 
to English:
\end{verbatim}

\section{Identifying parties from record}

\label{Appendix Parties}
\begin{figure}[h]
    \centering
    \includegraphics[width=\linewidth]{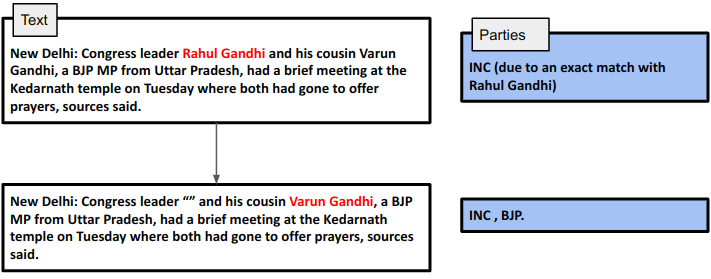}
    \caption{An example of how parties are extracted for a particular article.}
    \label{fig:party extraction}
\end{figure}

\textcolor{black}{We curated a dataset containing an entry of popular politicians of each party and the acronym used to refer to the party. We enhanced this list by including alternate names for prominent politicians, such as Maya (Mayawati) and NaMo (Narendra Modi), to improve matching accuracy. We then sort the entries based on length and start matching from the largest entry to the smallest entry, replacing each time an exact match has been found. This accounts for the fact that sometimes the keywords may be nested, and just an exact match could be erroneous. Table \ref{tab:parties dataset} shows a sample of the parties dataset. A pictorial example can be seen in Figure \ref{fig:party extraction}.}

\begin{table}[htbp]
    \caption{A sample of the parties dataset used to identify the political parties mentioned in a news headline.}
    \small
    \centering
    \setlength{\tabcolsep}{34pt}
    \begin{tabular}{|c|c|} 
        \hline 
        \textbf{INC} & \textbf{BJP} \\ 
        \hline 
        Priyanka Gandhi & Narendra Modi \\  
        Rahul Gandhi & Varun Gandhi \\ 
        Sonia Gandhi & Amit Shah \\ 
        Gandhi & Modi \\ 
        \hline
    \end{tabular}
    \label{tab:parties dataset}
\end{table}

\section{Keywords used for rallies}
\label{Appendix Rally Keywords}
The full list of speech event terms used to identify rallies: \\
\textit{speech, address, talk, oration, lecture, presentation, discourse, monologue, sermon, declaration, statement, pronouncement, elocution, rally, event, assembly, gathering, meeting, convocation, conclave, conference, convention, congregation, protest, demonstration, parade, march, road show, roadshow, campaign}

\section{Masking parties and politicians}
\label{Appendix Masking}
We also ran experiments to check if the classifier is robust in understanding the context properly surrounding conflict frames and if masking the politicians (with \texttt{<PERSON>}) and party (with \texttt{<PARTY>}) changes the metrics. No discernible deviations were observed in the metrics, suggesting that masking does not significantly impact the classifier's outcomes. Table \ref{Other_Metrics} provides an overview of the performance metrics associated with this masked approach. 

\begin{table}[htbp]
    \caption{Performance Metrics for a classifier trained to predict frames with a masked dataset.}
    \small
    \centering
    \setlength{\tabcolsep}{13pt}
    \begin{tabular}{|l|c|c|c|}
        \hline
        & \textbf{Precision} & \textbf{Recall} & \textbf{F1-Score} \\
        \hline
        Conflict & 0.86 & 0.92 & 0.89 \\
        Non-Conflict & 0.84 & 0.73 & 0.77 \\
        Macro average & 0.85 & 0.82 & 0.83 \\
        \hline
    \end{tabular}
    \label{Other_Metrics}
\end{table}

\section{Hyperparameters}
Our hyperparameter search was completed using a random selection method. Our objective was to maximize the average macro average F1-score. In this experiment, we adjusted several parameters. For class weights related to conflict and non-conflict classes, we applied a uniform distribution ranging from 1 to 10. The learning rate was set with a uniform distribution between $1 \times 10^{-6}$ and $1 \times 10^{-3}$, while the number of epochs followed an integer uniform distribution from 1 to 20. The final hyperparameters can be seen in table \ref{tab:Hyperparameters}.

\label{Appendix Hyperparameters}
\begin{table}[htbp]
    \caption{Hyperparameters for the classifier used to identify frames from election-related news headlines.}
    \small
    \centering
    \setlength{\tabcolsep}{28pt}
    \begin{tabular}{|l|c|}
        \hline
        \textbf{Hyperparameter} & \textbf{Value} \\
        \hline
        Conflict Class Weight & 1.69 \\
        Non-Conflict Class Weight & 9.01 \\
        Learning Rate & $6.008 \times 10^{-5}$ \\
        Epochs & 9 \\
        \hline
    \end{tabular}
    \label{tab:Hyperparameters}
\end{table}

\section{Identification of aggressors and targets}
\label{Appendix Agrressors}
The attached prompt (below) was used to identify the aggressor(s) and target(s) from a conflict-frame headline.

\begin{table}[htbp]
    \caption{Metrics for the identification of aggressors and targets from a headline.}
    \small
    \centering
    \setlength{\tabcolsep}{15pt}
    \begin{tabular}{|c|c|c|c|}
        \hline
        \textbf{Class} & \textbf{Precision} & \textbf{Recall} & \textbf{F1-Score} \\
        \hline
        Aggressor & 0.99 & 1.0 & 0.995 \\
        Target & 0.98 & 1.0 & 0.99 \\
        Overall & 0.985 & 1.0 & 0.993 \\
        \hline
    \end{tabular}
    \label{GPT Metrics}
\end{table}

\textcolor{black}{The classification report for the identification of aggressors and targets can be seen in table \ref{GPT Metrics}. A true positive is a headline for which the aggressor is identified correctly. A false negative is a headline for which the aggressor is not identified. A false positive is a headline for which the aggressor is identified incorrectly. The same definitions were used for identifying the metrics for the target.}

\small
\begin{verbatim}
Given a set of headlines, identify the 
speaker/subject (some political entity 
/ entities) and the target/object (some,
maybe different or similar political 
entity / entities) for each headline.

I will give you a few examples and 
the example output format. You 
can use this format to provide 
the output for the headlines 
I give you.

The output format is:
{
 "headline": "headline text",
 "speaker": [list of political entities],
 "target": [list of political entities]},
...
}
If there is no speaker or target, you can 
leave the list empty. 

Some examples:
Headline 1: 10% quota to Economically 
Weaker: It’s too late, Modi should 
have done it in 2014, says state 
Congress leader
Speaker: Congress
Target: Modi
Output:
{
  "headline": "10% quota to Economically 
    Weaker: It’s too late, Modi should 
    have done it in 2014, says state 
    Congress leader",
  "speaker": ["Congress"],
  "target": ["Modi"]
}
Headline 2: Narendra Modi must apologise 
for 'liquor' slur: Congress
Speaker: Congress
Target: Narendra Modi
Output:
{
  "headline": "Narendra Modi must 
  apologise for 'liquor' slur: Congress",
  "speaker": ["Congress"],
  "target": ["Narendra Modi"]
}
Headline 3: Maharashtra: Congress, 
NCP call Governor ‘pro-RSS’, 
boycott address
Speaker: ["Congress", "NCP"]
Target: RSS
Output:
{
  "headline": "Maharashtra: Congress, 
  NCP call Governor ‘pro-RSS’, 
  boycott address",
  "speaker": ["Congress", "NCP"],
  "target": ["RSS"]
}
In this case, notice the multiple 
speakers and the target is an indirect 
entity (RSS). Return the same format 
for all the headlines given below.
Now, here are the headlines:
\end{verbatim}

\section{(Over)emphasizing conflict frames}

Table \ref{tab:issues_table} shows the number of mentions of each issue and the proportion of matched articles mentioning the issue. Note, that there were no occurrences of any speeches mentioning the Ram Mandir. This could potentially be attributed to the limited number of keywords identified for that particular issue.
\begin{table}[htbp]
    \caption{For each issue: Total speech count and percentage of speeches addressing the issue}

    \centering

    \begin{tabular}{|l|c|c|}
        \hline
        \textbf{Issue} & \textbf{Occurrences} & \textbf{Proportion} \\
        \hline
        Opposition & 126 & 0.88 \\
        Economic Growth & 122 & 0.38 \\
        Corruption & 49 & 0.37 \\
        Farmers Issues & 76 & 0.32 \\
        Reservation & 52 & 0.30 \\
        Poverty & 92 & 0.28 \\
        Industrialisation & 79 & 0.22 \\
        Terrorism & 77 & 0.17 \\
        Unemployment & 50 & 0.15 \\
        Highways / Roads & 99 & 0.12 \\
        Education & 92 & 0.09 \\
        Healthcare & 89 & 0.09 \\
        CAA/NRC & 8 & 0.08 \\
        Inflation & 55 & 0.05 \\
        Water & 71 & 0.04 \\
        Electricity & 106 & 0.04 \\
        Women's Safety & 31 & 0.01 \\
        Cow Protection & 2 & 0.00 \\
        Ram Mandir & -- & 0.00 \\
        \hline
    \end{tabular}
    \label{tab:issues_table}
\end{table}

\section{Keywords for issues}
\label{Appendix Keywords}
The keywords used to match speeches to articles for the issues are seen in Table \ref{tab:categories_keywords}. Note, there is no entry for the issue \textit{Opposition} because the party matching algorithm was used to detect the presence of an opposition party.

\section{Confidence intervals and proportion percentages for coalitions as aggressors and targets}
We print the confidence intervals of headlines with the coalition as the aggressor in Table \ref{tab:alliance_attacked} and the confidence intervals with the coalition as the target in Table \ref{tab:alliance_being_attacked}.

\begin{table*}[t]
\caption{The list of issues and corresponding keywords used to identify these issues.}

\setlength{\tabcolsep}{4pt}
\begin{tabular}{|p{0.25\textwidth}|p{0.72\textwidth}|}
\hline
\textbf{Issue} & \textbf{Keywords} \\
\hline
Corruption & \textit{bail, black, bribe, cbi, chor, conspiraci, corrupt, croni, demonet, demonetis, expos, helicopt, investig, jail, jumla, lokpal, loot, nirav, pocket, probe, prosecut, rafal, raid, scam, scandal, steal, theft, thief} \\
\hline
Inflation & \textit{inflat, loan} \\
\hline
Unemployment & \textit{employ, job, labor, labour, mgnrega, unemploy} \\
\hline
Economic Growth & \textit{econom, economi, export, gdp, global, grow, growth, gst, job, money, prosper, scheme} \\
\hline
Farmers Issues & \textit{agrarian, agricultur, crop, farm, farmer, field, irrig, kisan, msp, price, seed, sugarcane} \\
\hline
Water & \textit{drink, drought, flood, irrig, pollut, river, water} \\
\hline
Highways / Roads & \textit{airport, buse, highway, road, traffic} \\
\hline
Electricity & \textit{electr} \\
\hline
Education & \textit{aiim, colleg, educ, iit, scientist, student, teacher, univers} \\
\hline
Poverty & \textit{basic, poor, poorest, poverti, rural} \\
\hline
Women's Safety & \textit{beti, girl, rape} \\
\hline
Terrorism & \textit{airstrik, blast, milit, pulwama, terrorist, uri} \\
\hline
Cow Protection & \textit{cow, slaughter} \\
\hline
Ram Mandir & \textit{ayodhya, masjid} \\
\hline
Reservation & \textit{adivasi, ambedkar, bhim, cast, casteism, class, dalit, discrimin, jat, obc, quota, section, tribal, tribe, upper} \\
\hline
Industrialisation & \textit{built, factori, forest, industri, invest, manufactur, project, sector, trade, trader} \\
\hline
Healthcare & \textit{ayushman, basic, doctor, health, healthcar, hospit, leprosi, medic, toilet} \\
\hline
CAA/NRC & \textit{amend, articl, citizenship, migrant, nrc, refuge} \\
\hline
\end{tabular}
\label{tab:categories_keywords}
\end{table*}

\begin{figure*}[htbp]
    \centering
    \begin{minipage}{0.48\textwidth}
        \centering
        \captionof{table}{Proportion of headlines with coalition as aggressor}
        \small
        \resizebox{\columnwidth}{!}{%
        \begin{tabular}{|l|c|l|c|c|c|}
            \hline
            Source & Year & Coalition & Lower & Upper & Est. \\
            \hline
            The Hindu & 2014 & UPA & .252 & .331 & .292 \\
            & & NDA & .303 & .386 & .345 \\
            \cline{2-6}
            & 2019 & UPA & .307 & .372 & .340 \\
            & & NDA & .257 & .320 & .289 \\
            \hline
            NDTV & 2014 & UPA & .310 & .382 & .346 \\
            & & NDA & .324 & .397 & .361 \\
            \cline{2-6}
            & 2019 & UPA & .325 & .378 & .352 \\
            & & NDA & .351 & .406 & .379 \\
            \hline
            The Indian Express & 2014 & UPA & .333 & .399 & .366 \\
            & & NDA & .318 & .383 & .351 \\
            \cline{2-6}
            & 2019 & UPA & .335 & .383 & .359 \\
            & & NDA & .319 & .367 & .343 \\
            \hline
            The Times of India & 2014 & UPA & .285 & .361 & .323 \\
            & & NDA & .287 & .363 & .325 \\
            \cline{2-6}
            & 2019 & UPA & .321 & .381 & .351 \\
            & & NDA & .303 & .362 & .333 \\
            \hline
            India Today & 2014 & UPA & .294 & .369 & .332 \\
            & & NDA & .279 & .352 & .316 \\
            \cline{2-6}
            & 2019 & UPA & .329 & .382 & .356 \\
            & & NDA & .317 & .371 & .344 \\
            \hline
            Zee News & 2014 & UPA & .183 & .359 & .271 \\
            & & NDA & .281 & .466 & .374 \\
            \cline{2-6}
            & 2019 & UPA & .281 & .331 & .306 \\
            & & NDA & .367 & .420 & .394 \\
            \hline
            Republic World & 2019 & UPA & .273 & .338 & .306 \\
            & & NDA & .337 & .405 & .371 \\
            \hline
        \end{tabular}}
        \label{tab:alliance_attacked}
    \end{minipage}
    \hfill
    \begin{minipage}{0.48\textwidth}
        \centering
        \captionof{table}{Proportion of headlines with coalition as target}
        \small
        \resizebox{\columnwidth}{!}{%
        \begin{tabular}{|l|c|l|c|c|c|}
            \hline
            Source & Year & Coalition & Lower & Upper & Est. \\
            \hline
            The Hindu & 2014 & UPA & .337 & .422 & .380 \\
            & & NDA & .445 & .532 & .489 \\
            \cline{2-6}
            & 2019 & UPA & .290 & .354 & .322 \\
            & & NDA & .485 & .553 & .519 \\
            \hline
            NDTV & 2014 & UPA & .369 & .443 & .406 \\
            & & NDA & .490 & .566 & .528 \\
            \cline{2-6}
            & 2019 & UPA & .369 & .424 & .397 \\
            & & NDA & .444 & .500 & .472 \\
            \hline
            The Indian Express & 2014 & UPA & .347 & .413 & .380 \\
            & & NDA & .468 & .536 & .502 \\
            \cline{2-6}
            & 2019 & UPA & .322 & .370 & .346 \\
            & & NDA & .484 & .535 & .510 \\
            \hline
            The Times of India & 2014 & UPA & .305 & .381 & .343 \\
            & & NDA & .464 & .544 & .504 \\
            \cline{2-6}
            & 2019 & UPA & .346 & .407 & .377 \\
            & & NDA & .431 & .493 & .462 \\
            \hline
            India Today & 2014 & UPA & .327 & .403 & .365 \\
            & & NDA & .517 & .595 & .556 \\
            \cline{2-6}
            & 2019 & UPA & .328 & .382 & .355 \\
            & & NDA & .518 & .574 & .546 \\
            \hline
            Zee News & 2014 & UPA & .378 & .573 & .476 \\
            & & NDA & .359 & .554 & .457 \\
            \cline{2-6}
            & 2019 & UPA & .330 & .382 & .356 \\
            & & NDA & .474 & .529 & .502 \\
            \hline
            Republic World & 2019 & UPA & .446 & .517 & .482 \\
            & & NDA & .410 & .480 & .445 \\
            \hline
        \end{tabular}}
        \label{tab:alliance_being_attacked}
    \end{minipage}
\end{figure*}

\end{document}